\documentclass[aps,twocolumn,eqsecnum,showkeys,aps,superscriptaddress, 6pt, pre ]{revtex4}
\usepackage{graphicx}
\usepackage{dcolumn}
\usepackage{xspace}
\usepackage{color}

\begin{document}

\title{Improved analysis of rare-earth magnetic superconductors}

\author{Preeti Suman Dash$^1$ and Salila Das}
\affiliation{Department of Physics, Berhampur University,
Berhampur 760007, Odisha, India}
\email{sd.phy@buodisha.edu.in}
\author{Prabir K. Mukherjee}
\affiliation{Department of Physics, Government College of Engineering
and Textile Technology, 12 William Carey Road, Serampore, Hooghly-712201,
India}
\email{pkmuk1966@gmail.com}
\date{\today}
\begin{abstract}
We present an improved analysis of the phase transitions in rare earth superconductor 
within Ginzburg-Landau theory. Our work is based on the systematic study of critical field and superconducting
order parameters in the presence of localized magnet due to rare earth atom.
We present the different phases
that can occur
and analyze the conditions of phase transition from the normal phase to the 
coexistence phase
of antiferromagnetism and superconductivity. We calculate the critical field and Ginzburg Landau parameters to
show the coexistence property. 
We compare our theoretical results with existing experimental results.
\end{abstract}
\maketitle
\newpage
\section{Introduction}

The magnetic and superconducting order parameters of rare earth compounds $(RNi_2B_2C)$ are in the same range \cite{muller2008rare,gupta2006superconductivity}
and thus attract the attention of researchers all around the globe. Antiferromagnetic and superconducting properties exhibited by these
compounds contradict each other. The superconducting behaviour is observed for both types of compounds, i.e., having nonmagnetic rare-earth elements  (Y, Lu) \cite{muller2008rare,nagarajan1994bulk} and elements (Tm, Er, Ho and Dy) with high magnetic moments \cite{lynn1997magnetic,gupta2006superconductivity}. Several compounds of the "$RNi_2B_2C$" group exhibit superconductivity, magnetism, and coexisting properties of both \cite{cava1994superconductivity,cho1995onset,siegrist1994crystal,canfield1998new,canfield1997angular,shorikov2006band} for different R (rare earth). The coexisting property in these compounds is due to the localized rare-earth $4f$ magnetic moments coupling with the conduction electrons $Ni (3d$) via exchange interaction.
The band structure calculation shows that the Fermi energy bands consist of $Ni(3d)$ orbital along 
with B and C 2p orbitals, and the dominant contribution is from Ni orbitals \cite{pickett1994luni,mattheiss1994electronic,shorikov2006band}. Superconducting condensation is associated with itinerant conduction electrons \cite{shorikov2006band}, and anisotropic magnetic properties are related to localized 4f electrons \cite{maki2002anisotropic,thalmeier2004unconventional}. These compounds exhibit various commensurate and incommensurate magnetic structures. The experimental tunneling data shows that $RNi_2B_2C$ superconductors are weak-coupling BCS type. They are type II superconductors with a small coherence length of 50-100 \AA. The different theoretical models and experimental observations have been proposed to study various features, precisely the upper critical field , specific heat, using neutron diffraction, magnetization measurements, and resistive transition curves \cite{thalmeier2004unconventional,amici1999theoretical,das2016effect,dasstudy,detlefs1996magnetic,eisaki1994competition,rathnayaka1997transport,tomy1995superconductivity, peng1998large,jensen2007competing}. The upper critical field provides knowledge of coupling strength, exchange interaction, the electronic structure,
and anisotropy of the order parameter.

The purpose of this present work is to have an exhaustive understanding of the phase transition from the normal
state to the coexistence state, and the thermodynamic properties  of these rare earth compounds theoretically. The Ginzburg-Landau
approach \cite{peng1998large,jensen2007competing,doh1999phenomenological,chela1988phenomenological,yanase2011ginzburg} helps to study the coexistence of magnetism and superconductivity in the multi band picture.
These theories explain magnetic fluctuations and anomalous behaviour of the upper critical field $H_{c2}$ \cite{mattheiss1994electronic,jensen2007competing,doh1999phenomenological}. 
We have done a systematic study of critical field and other Ginsburg parameters within a single phenomenological model, including 
magnetism and superconductivity. We construct our phenomenological model of two order parameters \cite{doh1999phenomenological} corresponding to the two 
dominant correlations based on the experimental information. 
The next section discusses the phenomenological framework to calculate the critical field and other parameters for rare-earth compounds. 
 Section 3  compares  our theoretical results with existing experimental results, and the last section contains the concluding remarks.

\section{The Free energy}
Ginzburg-Landau theory \cite{jensen2007competing,doh1999phenomenological,chela1988phenomenological,yanase2011ginzburg}  is a phenomenological approach to describe phase transition using a unified 
theoretical approach. It can explain a variety of phenomena 
without going into the technical process.

The Ginzburg-Landau theory is nonlinear second order differential equation which couples the 
spatial variation of a magnetic field with a superconducting order parameter. 
 The Ginzburg-Landau theory simply postulates the existence of macroscopic quantum wave function
$\psi (\bf {r})$, which is equivalent to an order parameter. The transition from superconducting to the normal 
state is assumed to take place at the thermodynamic critical field. 
So the difference in free energy between the normal  and superconducting state is given by the magnetic 
field energy of the excluded flux. The anti ferromagnetic order  
of rare earth magnetic superconductor can be described by $M_a$ and $M_b$. Here $M_a$ and $M_b$
are the magnetic order of two identical inter penetrating lattices labelled by 
"a" and "b". Further, we 
assume $M_a\cong -M_b$. For the superconducting order parameter, we define 
$\psi_a (\bf {r})$ and $\psi_b (\bf {r})$ as two superconducting order 
parameters 
in the lattices ''a'' and ''b'' respectively.

Now the free energy density of rare earth  magnetic superconductor  
should be expressed in terms of an expansion of $\psi_a$, $\psi_b$, $M_a$ and 
$ M_b$
\begin{widetext}
\begin{eqnarray}
F & = & \int d^3r[F_n + a_1 \left| \psi_a \right| ^2+a_2\left| \psi_b \right| ^2+
\frac 12 b_1 \left| \psi_a \right| ^4+\frac 12b_2\left| \psi_b \right| ^4
+\alpha (M_a^2+M_b^2)+\frac 12\beta (M_a^4+M_b^4) \nonumber \\
&&+2\delta M_aM_b
+\gamma_1\left| \psi_a \right|^2(M_a^2+M_b^2) 
+\gamma_2\left| \psi_b \right|^2(M_a^2+M_b^2)+2\eta\psi_a^2\psi_b^2-\kappa_1\psi_a^*\psi_b \nonumber \\
&&+\frac {1}{2m_a}\left| (-i\hbar \nabla-\frac {2e\mathbf{A}}{c})\psi_a 
\right| ^2 
+\frac {1}{2m_b}\left| (-i\hbar \nabla-\frac {2e\mathbf{A}}{c})\psi_b 
\right| ^2 \nonumber\\
&&+\kappa_2(i\hbar \nabla-\frac {2e\mathbf{A}}{c})\psi_a^*
(-i\hbar \nabla-\frac {2e\mathbf{A}}{c})\psi_b+\frac {H^2}{8\pi}]
\label{free1}
\end{eqnarray}
\end{widetext}

where $F_n$ is the free energy density of the normal phase. $a_1$, $a_2$, $b_1$, 
$b_2$, $\alpha$, $\delta$ and $\beta$ are material 
parameters. $\gamma_1$, $\gamma_2$, $\eta$, $\kappa_1$ and $\kappa_2$ are the coupling 
constants. $\gamma_1$ and $\gamma_2$ are assumed to be 
positive for the stability of the superconducting state. $e$, $m_a$ and $m_b$ are the elementary electron charge and mass, 
respectively.  Here $\delta>0$, $\beta>0$ and $b>0$. The parameter $a_1$ 
is proportional to $(T-T_{C1})$, $a_2$ is proportional to $(T-T_{C2})$ 
and $\alpha$ is proportional to $(T-T_{af})$. 
Thus $a_1=a_{01}(T-T_{C1})$, $a_2=a_{02}(T-T_{C2})$, and $\alpha=\alpha_0(T-T_{af})$. $a_{01}$, $a_{02}$ and $\alpha_0$ are 
positive constants. The $\psi_a$ and $\psi_b$ are the superconducting order parameters associated with Ni(3d).

First, we consider the uniform system in zero field $H=0$. 
After the minimization of  Eq.(\ref{free1}) with respect to $\psi_a$, $\psi_b$,
$M_a$ and $M_b$, we get the following four stable phases:

\noindent
I) Normal phase (N): $|\psi_a|=0$, $|\psi_b|=0$, $M_a=0$, $M_b=0$.
This phase exists for $a_1>0$, $a_2>0$ and $\alpha>0$.\\
(II) Anti ferromagnetic phase (AFM): $|\psi_a|=0$, $|\psi_b|=0$, $M_a \neq 0$, $M_b \neq 0$.
This phase exists for $a_1>0$, $a_2>0$ and
$\alpha<0$.\\
(III) Superconducting phase (SC): $|\psi_a| \neq 0$,  $|\psi_b|\neq 0$, $M_a=0$, $M_b=0$.
This phase exists for $a_1<0$, $a_2<0$ and $\alpha>0$.\\
(III) Coexistence of superconductivity and antiferromagnetism phase (AFS):
 $|\psi_a| \neq 0$,  $|\psi_b|\neq 0$, $M_a \neq 0$, $M_b \neq 0$. This phase exists for $a_1<0$, $a_2<0$ and $\alpha<0$.

From these solutions, it is clear that six types of phase transition are 
possible : N-SC, N-AFM, N-AFS, SC-AFM, AFM-AFS, SC-AFS.
The N-SC, N-AFM and SC-AFM transitions are second order. 
We will now discuss the N-AFS phase transition. The spontaneous magnetization in the AFS 
phase ($M_a=-M_b$) is given by

\begin{equation}
M^2_{sa}=M^2_{sb}=\frac {(\delta-\alpha-\gamma_1\left| \psi_a \right|^2-
\gamma_2\left| \psi_b \right|^2)}{\beta}
\label{mag}
\end{equation}

Now the substitution of Eq.(\ref{mag}) into Eq.
(\ref{free1}), we get
\begin{widetext}
\begin{eqnarray}
F&=&\int d^3r[F_n^* + a_1^* \left| \psi_a \right| ^2+a_2^* \left| \psi_b \right| ^2+
\frac 12 b_1^* \left| \psi_b \right| ^4+\frac 12b_1^* \left| \psi_b \right| ^4
+2\eta^* \psi_a^2\psi_b^2-\kappa_1\psi_a^*\psi_b \nonumber \\
&&+\frac {1}{2m_a}\left| (-i\hbar \nabla-\frac {2e\mathbf{A}}{c})\psi_a
\right| ^2 
+\frac {1}{2m_b}\left| (-i\hbar \nabla-\frac {2e\mathbf{A}}{c})\psi_b
\right| ^2 \nonumber \\
&&+\kappa_2(+i\hbar \nabla-\frac {2e\mathbf{A}}{c})\psi_a^*
(-i\hbar \nabla-\frac {2e\mathbf{A}}{c})\psi_b
+\frac {H^2}{8\pi}]
\label{free2}
\end{eqnarray}
\end{widetext}
where 

$F_n^*=F_n-\frac {\alpha^{2}}{\beta}-\frac {\delta^2}{\beta}+\frac 
{\delta \alpha}{\beta}$, 

$a_1^*=a_1+\frac {2\delta \gamma_1}{\beta}-\frac {2\gamma_1\alpha}{\beta}$,

$a_2^*=a_2+\frac {2\delta \gamma_2}{\beta}-\frac {2\gamma_2\alpha}{\beta}$,

$b_1^*=b_1-\frac {2\gamma_1^2}{\beta}$,

$b_2^*=b_2-\frac {2\gamma_2^2}{\beta}$,

$\eta^*=\eta-\frac {\gamma_1\gamma_2}{\beta}$.

Minimization of Eq.(\ref{free2}) with respect to $\psi_a$ and $\psi_b$ 
(for uniform system and $\kappa_1=0$) yields

\begin{equation}
\left|\psi_a\right|^2=\frac {a_1^*b_2^*-2\eta^*a_2^*}{4\eta^{*2}-b_1^*b_2^*}
\label{wave1}
\end{equation}

\begin{equation}
\left|\psi_b\right|^2=\frac {a_2^*b_1^*-2\eta^*a_1^*}{4\eta^{*2}-b_1^*b_2^*}
\label{wave2}
\end{equation}

Substitution Eq. (\ref{wave1}) and Eq. (\ref{wave2}) into Eq. (\ref{mag}), we get
\begin{widetext}
\begin{equation}
M^2_{sa}=M^2_{sb}=\frac {(\delta-\alpha)(4\eta^{*2}-b_1^*b_2^*)
-\gamma_1(a_1^*b_2^*-2\eta^*a_2*)-\gamma_2(a_2^*b_1^*-2\eta^*a_1^*)}
{\beta(4\eta^{*2}-b_1^*b_2^*)}
\label{mag1}
\end{equation}
\end{widetext}
Equations (\ref{wave1}) (\ref{wave2}) and (\ref{mag1}) are the values of superconducting
order parameters and spontaneous magnetization in the AFS phase. 

$T_{C1}>T<T_{af}$ and $T_{C2}>T<T_{af}$ are the conditions for the existence of the AFS phase. 
Then $a_1$, $a_2$ and $\alpha$ are negative. For the AFS phase, both 
Eq.(\ref{wave1}), Eq. (\ref{wave2}) and 
Eq. (\ref{mag1}) must be positive. Thus the necessary conditions for the
existence of the AFS phase are 

(1)   $a_1^*b_2^*>2\eta^*a_2^*$,

(2) $a_2^*b_1^*>2\eta^*a_1^*$,

(3)  $4\eta^{*2}>b_1^*b_2^*$

(4) $(\delta-\alpha)(4\eta^{*2}-b_1^*b_2^*)>
\gamma_1(a_1^*b_2^*-2\eta^*a_2^*)+\gamma_2(a_2^*b_1^*-2\eta^*a_1^*)$.

The above four conditions should hold simultaneously for the existence of the AFS phase. 
In this case, the free energy will be in the lowest energy state. So the 
N-AFS phase transition occurs. The superconductor has three characteristic 
parameters associated with them : the Ginzburg Landau coherence length ($\xi$)
, the penetration depth ($\lambda$), and the Ginzburg Landau parameter ($\kappa$).
 
The correlation length, penetration depth, and critical magnetic field
can be calculated for the AFS phase by the same method as adapted for the normal superconductor which
are given by
\begin{equation}
\xi_{GLa}(T)=\xi_{GLa}(0)(T_{C1}^*-T)^{-1/2}
\label{corre1}
\end{equation}

\begin{equation}
\xi_{GLb}(T)=\xi_{GLb}(0)(T_{C2}^*-T)^{-1/2}
\label{corre2}
\end{equation}

\begin{equation}
\lambda_{GLa}(T)=\lambda_{GLa}(0)(T_{C1}^*-T)^{-1/2}
\label{pene1}
\end{equation}

\begin{equation}
\lambda_{GLb}(T)=\lambda_{GLb}(0)(T_{C2}^*-T)^{-1/2}
\label{pene2}
\end{equation}

\begin{equation}
H_{Ca}=\frac {2m_aa_{01}^*c}{\hbar e}(T_{C1}^*-T)
\label{critical1}
\end{equation}

\begin{equation}
H_{Cb}=\frac {2m_ba_{02}^*c}{\hbar e}(T_{C2}^*-T)
\label{critical2}
\end{equation}

where

$\xi_{GLa}(0)=\frac {\hbar}{\sqrt{2m_aa_{01}^*}} $,

$\xi_{GLb}(0)=\frac {\hbar}{\sqrt{2m_ba_{02}^*}} $,

$\lambda_{GLa}(0)=\sqrt{\frac{m_ac^2b_1^{**}}{8\pi e^2a_{01}^*}}$,

$\lambda_{GLb}(0)=\sqrt{\frac{m_bc^2b_2^{**}}{8\pi e^2a_{02}^*}}$,

$a_{01}^*=a_{01}-\frac {2\eta^*a_{02}}{b^2*}-\frac {2\gamma_1\alpha_0}
{\beta}+\frac {4\eta^*\gamma_2\alpha_0}{\beta b_2^*}$,

$a_{02}^*=a_{02}-\frac {2\eta^*a_{01}}{b^1*}-\frac {2\gamma_2\alpha_0}
{\beta}+\frac {4\eta^*\gamma_1\alpha_0}{\beta b_1^*}$,

$T_{C1}^*=\frac {a_{01}T_{C1}-\frac {2\eta^*a_{02}}{b_2^*}T_{C2}-\frac {2\gamma_1
\alpha_0}{\beta}T_{af}+\frac {2\gamma_2\eta^*\alpha_0}{\beta b_2^*}T_{af}
-\frac {2\delta\gamma_1}{\beta}+\frac {4\delta\gamma_2\eta^*}{\beta b_2^*}}{
a_{01}^*}$,

$T_{C2}^*=\frac {a_{02}T_{C2}-\frac {2\eta^*a_{01}}{b_1^*}T_{C1}-\frac {2\gamma_2
\alpha_0}{\beta}T_{af}+\frac {2\gamma_1\eta^*\alpha_0}{\beta b_1^*}T_{af}
-\frac {2\delta\gamma_2}{\beta}+\frac {4\delta\gamma_1\eta^*}{\beta b_1^*}}
{a_{02}^*}$,

$b_1^{**}=b_1^*-\frac {4\eta^{*2}}{b_2^*}$,

$b_2^{**}=b_2^*-\frac {4\eta^{*2}}{b_1^*}$.

\section{Results and discussion}

We have determined the value of the upper critical magnetic field for various 
rare earth compounds, specifically $HoNi_2B_2C$, $ErNi_2B_2C$, $TmNi_2B_2C$ and 
$DyNi_2B_2C$ using Eq. (\ref{critical1} and Eq. \ref{critical2}) from the previous section.The curves for both the equations are
more or less identical. The corresponding graphs are plotted along with experimental results in figures 1-4. The numerical results are given in TABLE I.

It is observed that the critical field plot for all the compounds, shows similar trends as that of the experimental one. The theoretical value of $Hc_2(0)$ for $HoNi_2B_2C$ is 5.9 T, which is larger than the experimental value of 0.84 T. Similarly, for $TmNi_2B_2C$  $H_{c2}(0)$ is 3.9 while the experimental value is 2.0. However, for $DyNi_2B_2C$ and $ErNi_2B_2C$, the theoretical values of $Hc_2(0)$ are 0.21 T and 0.78 T, respectively which matches their experimental ones, i.e., 0.7 T and 0.112-0.113 T. The coherence length and penetration depth plotted using the above equations for all the compounds increases with temperature and diverge as T tends to $T_c$. 

\section{Conclusion}

Ginzburg's theory is valid in the superconducting phase boundary. Our multi-band model describes the behaviour and mutual influence of two magnetic and two superconducting order parameters. We have
studied the coexistence of superconducting and antiferromagnetism for rare
earth superconductors using Ginzburg Landau theory. We have calculated the effect of coherence length, penetration depth and critical field with temperature using the above model. Our
theoretical results agree with the experimental observations. Our model
describes some of the physical properties of the system in the coexisting
state. We do not find any anomalous behavior in the measurements of
the upper critical field.

\newpage
\section{References}
\bibliographystyle{unsrt}
\bibliography{references}

\begin{widetext}
\newpage
\begin{figure}[ht]
\includegraphics[height=14.4cm,width=17.4cm,angle=0]{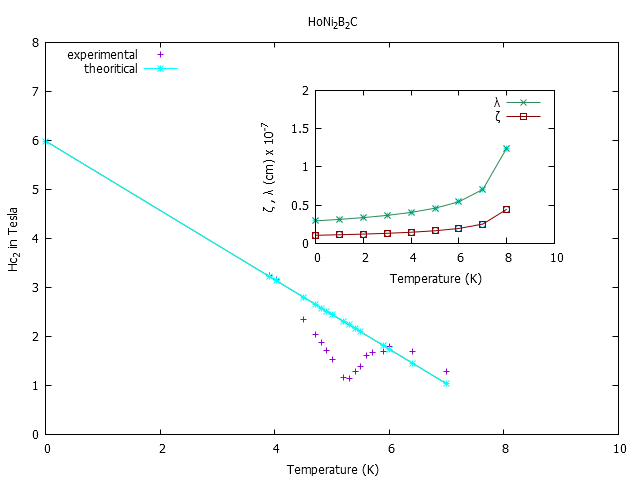}
\caption{Variation of upper critical magnetic field $Hc_2$ with the temperature for $HoNi_2B_2C$ \cite{alleno2010interplay}. Inset shows the Ginzburg-Landau parameters (coherence length and penetration depth) with temperature.}
\label{fig1}
\end{figure}

\newpage
\begin{figure}[ht]
\includegraphics[height=14.4cm,width=17.4cm,angle=0]{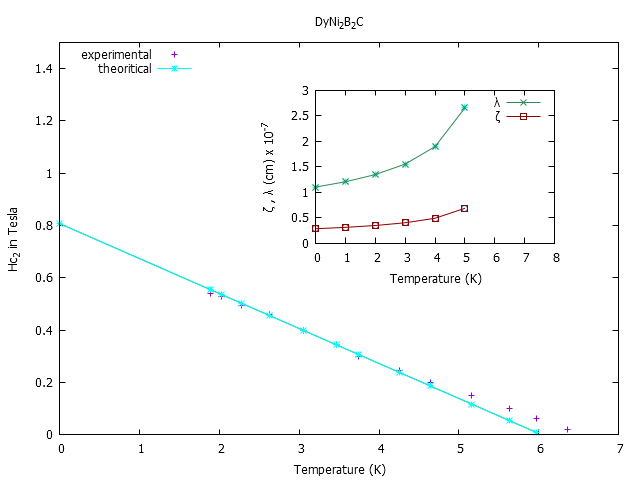}
\caption{The upper critical field $Hc_2$ as a function of temperature for $DyNi_2B_2C$ alongwith experimental results \cite{tomy1995superconductivity}. Inset shows Ginzburg-Landau parameters (coherence length and penetration depth) with temperature. The calculated $\zeta(0)$ 285.6 \AA                 
 while the experimental one is given by 220 \AA.}
\label{fig2}
\end{figure}

\newpage
\begin{figure}[ht]
\includegraphics[height=14.4cm,width=17.4cm,angle=0]{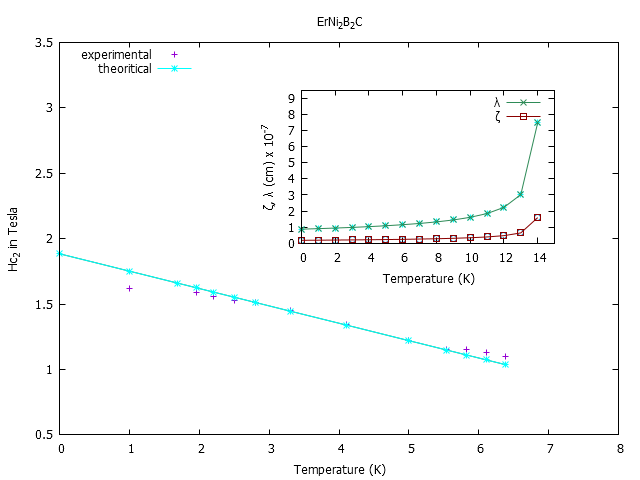}
\caption{The variation of $Hc_2(0)$ for $ErNi_2B_2C$ with temperature is plotted along with experimental values \cite{jensen2004magnetic}. Inset shows the variation of Ginzburg-Landau parameters (coherence length and penetration depth) with temperature.}
\label{fig3}
\end{figure}

\newpage
\begin{figure}[ht]
\includegraphics[height=14.4cm,width=17.4cm,angle=0]{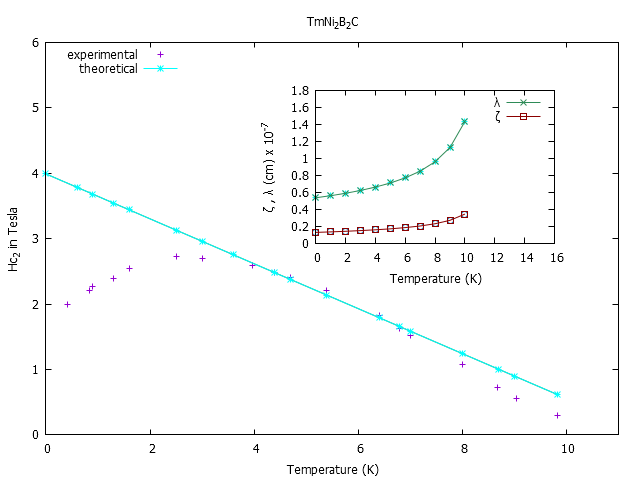}
\caption{The variation of  critical field of $TmNi_2B_2C$ with temperature along with experimental results \cite{naugle1999critical}. The inset shows the variation of Ginzburg-Landau parameters (coherence length and penetration depth) with temperature. The calculated $\zeta(0)$ 128.5 \AA while the experimental one is given by 180 \AA.},
\label{fig4}
\end{figure}

\newpage
\begin{table}
\caption{Values of different parameters for different rare earth magnetic superconductors}
\begin{ruledtabular}
\begin{tabular}{llllll}
$Compounds$ &$T_C$ in (K) & $T_N$ in (K) &$Hc_2(0)$ in Tesla &$\zeta_{GL}(0)$ in (\AA) &$\lambda_{GL}(0)$ in (\AA)\\
$HoNi_2B_2C$ &8.5 & 5.2 & 5.9 & 104.9 & 293.5\\
$DyNi_2B_2C$ &6.5 &10.5 &0.807 &285.6 &1102.5\\
$ErNi_2B_2C$ &11.5 & 6 & 1.9 & 186.9 & 877\\
$TmNi_2B_2C$ &11 &1.5 &3.99 &128.5 &535.4 \\
\end{tabular}
\end{ruledtabular}
\end{table}
\end{widetext}

\end{document}